\documentclass{article}
\usepackage[utf8]{inputenc}
\usepackage{geometry}
\newgeometry{hmargin={3cm,3cm}}   
\usepackage{color,amssymb,amsmath,epsfig,ifthen,graphicx,tabularx,xargs}
\usepackage{amsfonts}
\usepackage{amsthm}
\usepackage{algorithm}
\usepackage{algpseudocode}
\usepackage{enumitem,verbatim}
\usepackage{hyperref}
\hypersetup{
    colorlinks,
    linkcolor={red},
    citecolor={blue},
    urlcolor={blue!80!black}
}

\usepackage{cleveref}
\theoremstyle{remark}
\usepackage{mathrsfs}
\usepackage{stmaryrd}
\usepackage{caption}
\usepackage{subcaption}
\usepackage{pdflscape}
\usepackage{authblk}
\usepackage{natbib}
\usepackage{changepage}
\usepackage{booktabs}      
\usepackage{colortbl}      
\usepackage{array}         

\usepackage{mathtools}
\usepackage{xcolor}
\usepackage{multirow}
\usepackage{bm}
\usepackage[figuresright]{rotating}

\usepackage[textwidth=2cm,textsize=footnotesize]{todonotes}    

\def\bSig\mathbf{\Sigma}

\def\hat{\widehat}
\def\bX{{\bm X}}

\def\U{U_{jk} (\bar \bX_j, \bar \bX_k,  \bar \bX )}
\def\V{U_{jk} (\bar \bX_j, \bar \bX_k,  \bar \bX_{jk} )}
\def\zeta{\varphi}

\def\bbeta{{\bm \beta}}
\def\bSigma{{\bm \Sigma}}

\title{\textbf{Covariate Adjustment for Wilcoxon Two Sample Statistic and Test}}

\begin{document}

\author[1]{Zhilan Lou}
\author[2]{Jun Shao}
\author[3]{Ting Ye}
\author[4]{Tuo Wang}
\author[4]{Yanyao Yi}
\author[4]{Yu Du}

\affil[1]{School of Data Sciences, Zhejiang University of Finance and Economics, Hangzhou, Zhejiang, China}
\affil[2]{Department of Statistics, University of Wisconsin, Madison, Wisconsin, U.S.A.}
\affil[3]{Department of Biostatistics, University of Washington, Seattle, Washington, U.S.A.}
\affil[4]{Global Statistical Science, Eli Lilly and Company, Indianapolis, Indiana, U.S.A.\thanks{Corresponding author: \url{du_yu@lilly.com}}}

\maketitle

\begin{abstract}
We apply covariate adjustment to the Wincoxon two sample statistic and Wincoxon-Mann-Whitney test in comparing two treatments. 
The covariate adjustment through calibration not only improves efficiency in estimation/inference but also widens the application scope of  
the Wilcoxon  two sample statistic and Wincoxon-Mann-Whitney test to situations where covariate-adaptive randomization is used. We motivate how to adjust covariates to reduce variance, 
establish the asymptotic distribution of adjusted Wincoxon two sample statistic, and provide explicitly the guaranteed efficiency gain. 
The asymptotic distribution of adjusted  Wincoxon two sample statistic is invariant to all commonly used 
covariate-adaptive randomization schemes so that a unified formula can be used in inference regardless of which covariate-adaptive randomization is applied.

\end{abstract}

\section{Introduction}\label{sec: intro}

Consider a random sample  of $n$ units, each of which is assigned to one and only one treatment $j$ and results in an outcome $Y_j$ distributed with  an
unknown continuous  distribution $F_j$, $j=1,...,J $, where $J \geq 2$ is the number of treatments. An example  is a  clinical trial 
to study effects of $J$ medical products, in which units are typically patients. 
The simplest way of assigning treatments is 
simple randomization that assigns $n$ units completely at random with a pre-determined probability $\pi_j>0$ to treatment $j$, $\sum_{j=1}^J \pi_j =1$. 

Let $A \in \{1,...,J\}$  be the treatment assignment 
and $A_i$ be the treatment assignment for  unit $i$ in the random sample, $i=1,...,n$. Details of generating $A_i$'s  are given in Section 2. If $A_i=j$, then unit $i$ is assigned to treatment $j$ and the observed outcome from unit $i$ is denoted by $Y_{iA_i} = Y_{ij} \sim $ (distributed as) $Y_j$. Estimation and inference on unknown characteristics in $F_1,...,F_J$ can be carried out based on outcomes $Y_{iA_i}$, $i=1,...,n$. 

For comparing two fixed treatments $j$ and $k$,
 the well-known  Wilcoxon-Mann-Whitney rank-sum test \citep[pages 5-9]{Lehmann1975}
 is a valuable nonparametric alternative to the two sample t-test  (based on sample means of outcomes from two treatment groups) that is 
criticized when the outcome $Y_j$ or $Y_k$ is not normally distributed and/or has large variance. 
The  Wilcoxon-Mann-Whitney test statistic 
is given by 
\begin{equation}\label{wil}
n_jn_k U_{jk} + \frac{n_j(n_j+1)}{2}, \quad \qquad U_{jk} = \frac{1}{n_jn_k} \sum_{i: A_i=j} \, \sum_{i':A_i=k} 
I (Y_{ij} \leq Y_{i'k}), 
\end{equation}
 focusing on the number of outcome pairs $Y_{ij}$ and $Y_{i'k}$ with $Y_{ij} \leq Y_{i'k}$, 
where $I(B)$ is the indicator of event $B$,  $n_t$ is the number of $i$'s with $A_i=t$, and $t=j, k$. The $U_{jk}$ in (\ref{wil}) is  called the Wilcoxon two sample statistic \citep[page 175]{Serfling}, a special case of  the two sample U-statistic 
for estimating the treatment effect $\theta_{jk}= E(U_{jk}) = P ( Y_{j}\leq Y_{k})$. 
  Under the null hypothesis $H_0 \! : F_j = F_k$ (i.e., there is no difference in populations under treatments $j$ and $k$), $\theta_{jk} = 1/2$ and, thus, the Wilcoxon-Mann-Whitney test rejects $H_0$ when  $  U_{jk} $ is far way from $1/2$. More details are given in Section 3.1.

In many studies there exists  covariate information related with the  outcomes, useful for gaining estimation efficiency.
In clinical trials, for example, there are baseline covariates not affected by treatments,
such as  patient's age, sex, geographical location, occupation, education level, disease stage, etc. 
Utilizing covariates to improve efficiency of estimation and inference is referred to as covariate adjustment. 

The Wilcoxon two sample statistic $U_{jk}$ in (\ref{wil}) 
does not make use of any covariate and, thus, it may be improved by covariate adjustment.  If a correct model between outcomes and covariates can be specified, then  $U_{jk}$  can be improved  through model fitting with covariates. 
However, such a model-based approach relies heavily on the model correctness.
Because a correct model may not be easily specified in applications, 
model-free approaches for covariate adjustment have caught on recently. In the regulatory  agencies of clinical trials, for example, it is particularly recommended to utilize covariates ``under approximately the same minimal statistical assumptions that would be needed for unadjusted estimation''
\citep{ICHE9,ema:2015aa,fda:2019aa}. 
Note that the consistency and asymptotic normality of Wilcoxon two sample $U_{jk}$ in (\ref{wil}) is established under no assumption other than simple randomization and $n\to \infty$ \citep{Jiang2010}. 

The purpose of this paper is to apply covariate adjustment to $U_{jk}$ in (\ref{wil}) and the related Wilcoxon-Mann-Whitney test,  
 through  model-free covariate calibration  considered as early as in 
\cite{Cassel:1976aa} for survey problems and well summarized in  \cite{Sarndal:2003aa}. 
The covariate calibration has been shown to be effective in gaining efficiency for functions of sample means or  estimators  from generalized estimation equations
\citep[among others]{Yang:2001aa,Freedman:2008ab,zhang2008improving, Moore:2009tu,Lin:2013aa, Vermeulen2015,Wang:2019aa, Liu:2020aa,Benkeser2021, ZhangM2021, cohen2021noharm,Wang:2021wg, Ye2021better, Bannick}. But our covariate adjustment for $U_{jk}$ in (\ref{wil}) is created  by directly using the covariance between $U_{jk}$ and adjusted covariates without any assumption. It guarantees an asymptotic efficiency gain over the unadjusted $U_{jk}$ and provides an invariant inference formula for all 
commonly used  covariate-adaptive randomization schemes (Section 2.1), including simple randomization.
It also gains additional efficiency in comparing two treatments when there are more than two treatments ($J > 2$).

After introducing notation and randomization for treatment assignments, 
in Section 2.2 we derive covariate adjustment/calibration  for the Wilcoxon two sample statistic $U_{jk}$ in (\ref{wil}), motivated by why this adjustment guarantees efficiency gain. 
In Section 2.3 we establish that the proposed covariate adjusted  statistic 
is  asymptotically normal under all commonly used covariate-adaptive randomization, without any  assumption.  
The covariate adjusted statistic is guaranteed to be more efficient than the unadjusted $U_{jk}$ with  explicitly given efficiency gain. Our asymptotic result for  adjusted Wilcoxon two sample statistic  is 
invariant to randomization schemes for treatment assignments if covariates used in randomization are included in calibration, which is an advantage of our adjustment since practitioners can use 
 a unified formula  for inference, regardless of which covariate-adaptive  randomization is used. 
This unified formula property does not hold for unadjusted $U_{jk}$ as well as the sample means for t-tests, since their asymptotic distributions under covariate-adaptive randomization  are different 
from those under simple randomization. 
Based on the asymptotic theory and variance estimation, in Section 3 we propose a covariate adjusted 
Wilcoxon-Mann-Whitney test for $H_0: F_j = F_k$, which is more powerful (in terms of Pitman's asymptotic relative efficiency) than the two sample t-test and  
unadjusted Wilcoxon-Mann-Whitney test when covariates are useful, regardless of the type of outcome distribution. Even for normally distributed outcomes under which the two sample t-test is more powerful  than the unadjusted Wilcoxon-Mann-Whitney test, the adjusted  Wilcoxon-Mann-Whitney test may have an additional improvement and better than the two sample t-test. Furthermore, 
the Wilcoxon statistic in (\ref{wil})  and its adjustment are invariant to any monotone  transformation of outcome, as it is a function of outcome ranks, unlike the two sample t-test for which a serious effort may be needed to  find a suitable transformation such as the Box-Cox transformation
when $Y_j$ appears to be non-normal (see Section 5 for a real data example).  
Section 4 contains  some simulation results to  examine finite performances and complement the asymptotic theory.
In Section 5, a real data example is considered for illustration.

\section{Calibrating Wilcoxon Two Sample Statistic}

Following the notation in Section 1, observed outcomes are $Y_{iA_i}$, $i=1,...,n$, the number of outcomes under treatment $j$ is $n_j$, $\sum_{j=1}^J n_j = n$, 
and the Wilcoxon two sample statistic  to compare treatments $j$ and $k$ is $U_{jk}$ in (\ref{wil}) unbiased for  $\theta_{jk}$, $j=1,...,J$, $ k=1,..., J$, $j \neq k$. 

\subsection{Covariate-adaptive randomization of treatment assignments} 

Covariate adjustment can be started 
at the stage of treatment assignment, i.e., generating treatment assignments $A_1,...,A_n$, prior to obtaining outcomes.  
Simple randomization  assigns  treatments completely at random with probability $P(A_i =j ) = \pi_j$ to assign unit $i$ to treatment $j$
with pre-specified $\pi_j>0$, $\sum_{j=1}^J \pi_j = 1$. Simple randomization 
does not make use of covariates and  may yield  treatment proportions that substantially deviate from the target $\pi_j$ across levels of some baseline prognostic factors, especially when units are sequentially arrived.  
To balance the number of units in each treatment group across baseline prognostic factors,  covariate-adaptive randomization has become the new norm.
From 1989 to 2008, covariate-adaptive randomization was used in more than 500 clinical trials \citep{Taves2010};
among nearly 300  trials published in two years, 2009 and 2014,
237 of them applied covariate-adaptive randomization \citep{Ciolino:2019aa}.
The two most popular covariate-adaptive randomization schemes are the stratified permuted block \citep{Zelen:1974aa} and Pocock-Simon's minimization \citep{Taves:1974aa, Pocock:1975aa}. Other schemes and details of each can be found in two reviews \cite{Schulz:2018aa} and \cite{Shao:2021aa}.

Because treatment assignments are more balanced under covariate-adaptive randomization, the resulting estimators are more efficient compared with that under simple randomization. However,  covariate-adaptive randomization generates a dependent sequence of treatment assignments $A_1,...,A_n$ and, thus,  conventional method under simple randomization to derive asymptotic properties of estimators  is not applicable \citep{ema:2015aa,fda:2019aa}.  
For example, under the stratified permuted block randomization,
the Wilcoxon two sample statistic $U_{jk}$ in (\ref{wil}) has an asymptotic distribution different from that under simple randomization. Thus, to use the Wilcoxon-Mann-Whitney test under the stratified permuted block randomization,
one has to derive its asymptotic distribution. 
This seems to be what we need to do but we remedy the issue by covariate adjustment/calibration. 

\subsection{Covariate calibration} 

Let $\bX$ be the vector of covariates used for adjustment and $\bX_i$ be its value from unit $i$. We assume that all baseline covariates used in covariate-adaptive randomization are included in $\bX$, which is important as we explained later in Section 3. 
We assume that  $\bX$ is  chosen so that  $\bSigma = {\rm Var}( \bX)$ is finite and non-singular. 
Also, $\bSigma$ does not depend on $j$ since  covariates are not affected by treatments, which is the reason why we can gain efficiency over $ U_{jk}$ in (\ref{wil}) by  covariate adjustment. 

The idea of covariate calibration is to add  ``estimators" of zero to $U_{jk}$ in (\ref{wil}) to form 
\begin{equation}
 \U = U_{jk} +  (\bar{\bX}_{j}-\bar{\bX})^\top  \bbeta_j
 - (\bar{\bX}_{k}-\bar{\bX})^\top  \bbeta_k , 
 \label{adj}
\end{equation}
where 
$\bar \bX_{t}$ is the sample mean of $\bX_i$'s in treatment group 
$t=j$ or $k$,  $\bar \bX$ is the sample mean of all $\bX_i$'s, 
${\bm a}^\top$ is the transpose of ${\bm a}$, and $\bbeta_j$ and $\bbeta_k$ are nonrandom vectors chosen to reduce the variance of $U_{jk}$ and are estimated later if they depend on unknown quantities. 
The covariate calibration refers to calibrate $\bar\bX_{t}$ by $\bar\bX$ with covariate data not in treatment group $t$. 
  
To find out what $\bbeta_j$ and $\bbeta_k$ reduce $ {\rm Var} (U_{jk})$, we calculate the variance  of $ \U  $ in (\ref{adj}) under simple randomization. 
By the independence and exchangeability  of data  and  the independence of data in different treatment groups under simple randomization,  
\begin{align*}
{\rm Cov} ( U_{jk}, \,  \bar \bX_{j}  ) & = (\pi_jn)^{-1}{\rm Cov} \big\{   I(Y_{ij}\leq Y_{i'k}), \,  \bX_{ij}  \big\} +o(n^{-1})\\
& = (\pi_jn)^{-1}{\rm Cov} \big\{ E [ I(Y_{ij}\leq Y_{i'k}) \mid Y_{ij} , \bX_{ij} ], \,  \bX_{ij}  \big\} +o(n^{-1})\\
& = (\pi_jn)^{-1}{\rm Cov} \big\{  1-F_k(Y_{ij}), \,  \bX_{ij}    \big\} +o(n^{-1})\\
& = - \, (\pi_jn)^{-1} {\bm C}_{jk} +o(n^{-1}),
\end{align*}
where $
{\bm C}_{jk} =  {\rm Cov} \{  F_k(Y_{j}) , \, \bX_{j}\}$, 
$\bX_{ij}$ (or $\bX_{j}$) is $\bX$ associated with $Y_{ij}$ (or $Y_{j}$), $o(n^{-1}) = n^{-1} o(1)$, and $o(1)$  denotes a term $ \to 0$ as $n \to \infty$. 
Similarly, 
$
{\rm Cov} ( U_{jk}, \,  \bar \bX_{k}  ) =  (\pi_kn)^{-1} \, {\bm C}_{kj} + o(n^{-1}) $, 
where ${\bm C}_{kj} =  {\rm Cov} \{  F_j(Y_{k}) , \, \bX_{k}\}$. 
By the independence of data in different treatment groups and these results about covariances, 
\begin{align*}
{\rm Var} \{ \U\} = & \ {\rm Var} \big\{U_{jk} +   (\bar{\bX}_{j}-\bar{\bX})^\top  \bbeta_j
-  (\bar{\bX}_{k}-\bar{\bX})^\top  \bbeta_k \big\}\\
= & \ {\rm Var} (U_{jk}) + 
{\rm Var} \big\{   ( \bar \bX_{j} - \bar \bX)^\top \bbeta_j\big\}  + 
{\rm Var} \big\{  ( \bar \bX_{k} - \bar \bX)^\top \bbeta_k\big\} \\
 & \ + 2 {\rm Cov}\big\{ U_{jk}, \, ( \bar \bX_{j} - \bar \bX)^\top \bbeta_j\big\}  - 2 {\rm Cov}\big\{ U_{jk}, \, ( \bar \bX_{k} - \bar \bX)^\top \bbeta_k\big\} \\
& \ -2 {\rm Cov}\big\{ ( \bar \bX_{j} - \bar \bX)^\top \bbeta_j, \, 
( \bar \bX_{k} - \bar \bX)^\top \bbeta_k \big\} \\
= & \ {\rm Var} (U_{jk})     + {\textstyle \frac{1-\pi_j}{\pi_jn}}\bbeta_j^\top \bSigma \bbeta_j + {\textstyle \frac{1-\pi_k}{\pi_kn}}\bbeta_k^\top \bSigma \bbeta_k\\
& \ - {\textstyle \frac{2(1-\pi_j)}{\pi_jn} \bbeta_j^\top }{\bm C}_{jk}  - {\textstyle \frac{2}{n}} \bbeta_j^\top {\bm C}_{kj}  -
  {\textstyle \frac{2(1-\pi_k)}{\pi_kn} \bbeta_k^\top }{\bm C}_{kj}
 - {\textstyle \frac{2}{n}} \bbeta_k^\top {\bm C}_{jk} \\
 & \  + {\textstyle \frac{2}{n}} \bbeta_j^\top \bSigma \bbeta_k +o(n^{-1}) ,
	\end{align*}  
where we used 
$$ \bar \bX_{j} - \bar \bX = \frac{n-n_j}{n} \bar \bX_{j} - \frac{n_k}{n} \bar \bX_{k} - \frac{n-n_j-n_k}{n} \bar \bX_{-jk} \vspace{-2mm} $$
with $\bar \bX_{-jk}$ denoting the sample mean of $\bX_i$'s not in treatment groups $j$ and $k$, and 
\begin{align*}
& \ {\rm Cov}\big\{ ( \bar \bX_{j} - \bar \bX)^\top \bbeta_j, \, 
( \bar \bX_{k} - \bar \bX)^\top \bbeta_k \big\} \\
= & \ \bbeta_j^\top \{ {\rm Var}(\bar \bX) -{\rm Cov}(\bar \bX_{j} , \bar \bX) - {\rm Cov}(\bar \bX_{k} , \bar \bX)   \}\bbeta_k \\ 
= &  - n^{-1} \bbeta_j^\top \bSigma \bbeta_k .
\end{align*}
It turns out that if we choose 
\begin{equation}\label{beta}
\bbeta_j =  \bSigma^{-1} {\bm C}_{jk}  \qquad 
\mbox{and} \qquad \bbeta_k =  \bSigma^{-1} {\bm C}_{kj},  
\end{equation}
then
\begin{align*}
{\rm Var} \{ \U \} = & \ {\rm Var} (U_{jk}) - {\textstyle  \frac{1-\pi_j}{\pi_j n}} \bbeta_j^\top \bSigma \bbeta_j - {\textstyle  \frac{1-\pi_k}{\pi_k n}} \bbeta_k^\top \bSigma \bbeta_k 
- {\textstyle  \frac{2}{n}} \bbeta_j^\top \bSigma \bbeta_k + o(n^{-1}) \\
 = & \ {\rm Var} (U_{jk})  -  \frac{(\pi_j\bbeta_k+ \pi_k\bbeta_j)^\top \bSigma 
 	(\pi_j\bbeta_k+ \pi_k\bbeta_j)}{ \pi_j\pi_k (\pi_j+ \pi_k)n } \\
 & \
 - \frac{(1-\pi_j-\pi_k) (\bbeta_j-\bbeta_k)^\top \bSigma (\bbeta_j-\bbeta_k)}{(\pi_j+\pi_k)n} + o(n^{-1}) \\
  \leq & \ {\rm Var} (U_{jk}) + o(n^{-1}) 
 \end{align*}
with equality holds if and only if 
${\bm C} _{jk} = {\bm C} _{kj}  =0$ (i.e.,  $\bar \bX _{j}$ and $ \bar \bX_{k}  $ are uncorrelated with $U_{jk}$ so that $U_{jk}$ cannot be improved by using $\bar \bX_{j}$ and $\bar \bX_{k}$ in the adjustment).
 
 When $J >2$, although we compare just two treatments $j$ and $k$, the proposed adjustment still uses covariates from all $J$ treatments, instead of those just from treatments $j$ and $k$, because utilizing all $\bX_i$'s results in more efficient adjusted estimators. Specifically,  if we only use covariates in treatment groups $j$ and $k$ for adjustment,
 then  adjustment  (\ref{adj}) should be changed to 
 $$ \V =  U_{jk} +  (\bar{\bX}_{j}-\bar{\bX}_{jk})^\top  \bbeta_j
 - (\bar{\bX}_{k}-\bar{\bX}_{jk})^\top  \bbeta_k , $$
 where $\bar \bX_{jk}$ is the sample mean of $\bX_i$'s in treatment groups $j$ and $k$, 
and  the same calculation leads to 
 \begin{align*}
 {\rm Var} \{ \V \}= & \ {\rm Var} (U_{jk}) - {\textstyle  \frac{\pi_k}{\pi_j(\pi_j+\pi_k) n}} \bbeta_j^\top \bSigma \bbeta_j - {\textstyle  \frac{\pi_j}{\pi_k (\pi_j+\pi_k)n}} \bbeta_k^\top \bSigma \bbeta_k 
 - {\textstyle  \frac{2}{n}} \bbeta_j^\top \bSigma \bbeta_k + o(n^{-1}) \\
 >  & \ {\rm Var} (U_{jk}) - {\textstyle  \frac{1-\pi_j}{\pi_j n}} \bbeta_j^\top \bSigma \bbeta_j - {\textstyle  \frac{1-\pi_k}{\pi_k n}} \bbeta_k^\top \bSigma \bbeta_k 
 - {\textstyle  \frac{2}{n}} \bbeta_j^\top \bSigma \bbeta_k + o(n^{-1}) \\
= & \ {\rm Var} \{ \U \}+ o(n^{-1}) ,
 \end{align*}
 where the strict inequality follows from $\pi_k /(\pi_j+\pi_k) < 1- \pi_j$ for any $j$ and $k$ when $J >2$ (i.e., $j$ and $k$ are not the only treatments) and at least one of ${\bm C}_{jk}$ and ${\bm C}_{kj}$ is nonzero (otherwise $>$ should be replaced by $=$).

To finish our proposal for covariate adjustment, we just need to substitute $\bbeta_j$ and $\bbeta_k$ in (\ref{beta}) by  estimators. 
Form (\ref{beta}), $\bbeta_j$ and $\bbeta_k$ involve unknown  $\bSigma$, ${\bm C}_{jk}$,  and ${\bm C}_{kj}$. 
The matrix $\bSigma$ can be consistently estimated by $\hat \bSigma =$ the sample covariance matrix of all $\bX_i$'s. From how ${\bm C}_{jk}$ and ${\bm C}_{kj}$ are  derived, they can be 
consistently estimated respectively by 
$$
\hat {\bm C}_{jk} = \frac{1}{n_jn_k} \sum_{i: A_i=j}\, \sum_{i': A_{i'}=k} 
	I( Y_{i'k} \leq Y_{ij})(  \bX_{ij} - \bar \bX_j) $$ 
and 
$$ \hat {\bm C}_{kj} = \frac{1}{n_jn_k} \sum_{i: A_i=j} \, \sum_{i': A_{i'}=k} 
I(Y_{ij} \leq Y_{i'k})(  \bX_{i'k} - \bar \bX_k) .
$$
Thus, consistent estimators of $\bbeta_j$ and $\bbeta_k$ are 
$\hat\bbeta_j = \hat \bSigma^{-1} \hat  {\bm C}_{jk}$ and  $\hat\bbeta_k = \hat \bSigma^{-1} \hat  {\bm C}_{kj}$, respectively.

Our proposed adjusted Wilcoxon two sample statistic is then
\begin{equation}\label{CA}
U_{jk}^{ \rm C} = U_{jk}  +  (\bar{\bX}_{j}-\bar{\bX})^\top  \hat\bbeta_j
- (\bar{\bX}_{k}-\bar{\bX})^\top  \hat\bbeta_k . 
\end{equation}
By the consistency of $ \hat\bbeta_j$ and $ \hat\bbeta_k$  and the previous discussion, we conclude that 
$U_{jk}^{ \rm C} $ has a variance no larger than that of $U_{jk}$ when $n$ is large, under simple randomization. 

The proposed covariate adjusted  $U_{jk}^{ \rm C} $ in (\ref{CA}) also works when covariate-adaptive randomization is used. 
The asymptotic ($n\to \infty$) distribution of $U_{jk}^{ \rm C} $ and the variance reduction property of $U_{jk}^{ \rm C} $ are shown  next, under 
covariate-adaptive randomization.

\subsection{Asymptotic Theory}


In the following we derive the asymptotic distribution of adjusted $U_{jk}^{ \rm C}$  in (\ref{CA}) and show that $U_{jk}^{ \rm C}$ is asymptotically more efficient than $U_{jk}$  or equivalent to $U_{jk}$ when 
${\bm C}_{jk} = {\bm C}_{kj} =0$, for all commonly used covariate-adaptive randomization schemes satisfying  
the following minimum condition.  
\begin{description}
	\item (D) 	Covariate-adaptive randomization is carried out using a discrete baseline covariate ${\bm Z} $  with finitely many joint levels;	 conditioned on $({\bm Z}_1,...,{\bm Z}_n)$, treatment assignments, outcomes, and 
	covariates are independent, where ${\bm Z}_i$ is the value of ${\bm Z}$ for the $i$th unit; $P(A_i=j \mid {\bm Z}_1,...,{\bm Z}_n) = \pi_j$ for all $i$; and
	for every level $z$ of ${\bm Z}$, $n_{zj} /n_z \rightarrow  \pi_j$ in probability as $n\to \infty$, where 
	$n_{zj}$ is the number of
	units with ${\bm Z}_i=z$ and $A_i=j$, and $n_z = \sum_{j=1}^J n_{zj}$.
\end{description}

Condition (D) is satisfied for most popular covariate-adaptive randomization schemes, including the stratified permuted block and Pocock-Simon's minimization \citep{Baldi-Antognini:2015aa}, 
as well as simple randomization (with ${\bm Z} =$ constant) that is not covariate-adaptive. 
 \vspace{4mm} 
  
\noindent
{\bf Theorem 1}. Assume (D) for randomization to generate $A_1,...,A_n$ and assume that ${\bm Z}$ used in covariate-adaptive randomization is included in $\bX$ for adjustment. Then, as $n \to \infty$, 
$\sqrt{n} ( U_{jk}^{ \rm C}-\theta_{jk} )$ converges in distribution to the normal distribution with mean 0 and variance $\tau_{jk}+ \tau_{kj} - \zeta_{jk}$, where 
\begin{align*}
\tau_{jk}  & = \frac{{\rm Var} \{F_k(Y_{j})\}}{\pi_j } , \qquad 
\tau_{kj} =  \frac{{\rm Var}\{ F_j(Y_{k})\}}{\pi_k  } ,  \qquad \mbox{and}  \\
\zeta_{jk}  & =  \frac{(\pi_j\bbeta_k+ \pi_k\bbeta_j)^\top \bSigma 	(\pi_j\bbeta_k+ \pi_k\bbeta_j)}{ \pi_j\pi_k (\pi_j+ \pi_k) }  +  \frac{(1-\pi_j-\pi_k) (\bbeta_j-\bbeta_k)^\top \bSigma (\bbeta_j-\bbeta_k)}{\pi_j+\pi_k}. 
\end{align*} 

The proof of Theorem 1 is  in the  Appendix. Here we elaborate the result in Theorem 1 in three aspects. 
\begin{enumerate}
	\item[(i)] The result  in Theorem 1 is not only valid for all covariate-adaptive randomization schemes satisfying (D), but also invariant in the sense that the limiting variance $
	\tau_{jk}+\tau_{kj}-\zeta_{jk}$  in Theorem 1 is the same for all covariate-adaptive randomization satisfying (D), which means
	a unified procedure can be used in inference, desirable for practitioners as
	no tailored formula to each randomization scheme is needed. 
    To achieve this, we must ensure that ${\bm Z}$ used in covariate-adaptive randomization is included in $\bX$. Without this condition, the asymptotic distribution of $U_{jk}^{ \rm C}$ varies with randomization scheme and is different from that in Theorem 1. 
    \item[(ii)] Under simple randomization, for the unadjusted $U_{jk}$, 
    $\sqrt{n} (U_{jk} - \theta_{jk})$ is asymptotically  normal  with mean 0 and variance  $\tau_{jk}+ \tau_{kj}$ \citep[pages 380-382]{Jiang2010}.  This immediately shows that the adjusted 
    $U_{jk}^{\rm C}$ is asymptotically more efficient than the unadjusted $U_{jk}$ under simple randomization, unless $\zeta_{jk} =0$, i.e.,  $\bar \bX _{j}$ and $ \bar \bX_{k}  $ are uncorrelated with $U_{jk}$, in which case $U_{jk}^{\rm C}$ and $U_{jk}$ have the same asymptotic distribution. 
    When covariate-adaptive randomization is applied, however, the asymptotic distribution of unadjusted $U_{jk}$ varies with randomization scheme and is not available from the literature for some covariate-adaptive randomization such as Pocock-Simon's minimization. 
	\item[(iii)] When $J >2$, although we compare just two treatments $j$ and $k$, the proposed adjustment is asymptotically more efficient than covariate adjustment just using covariates from treatment groups $j$ and $k$, as discussed in Section 2.2.  
\end{enumerate}

\section{Inference}

The  asymptotic distribution of $U_{jk}^{ \rm C}$ in Theorem 1 is explicit and useful for statistical inference  on unknown $\theta_{jk}$ considered in this section. 

\subsection{Testing}

Consider testing null hypothesis $H_0\!: F_j = F_k$.
Under $H_0$,  both $ F_k (Y_{ij})$ and $F_j (Y_{ik}) $ are uniform  on the interval $[0,1]$ since $F_j$ and $F_k$ are continuous and, hence, $\theta_{jk}=  \frac{1}{2}$ and $ \tau_{jk}+\tau_{kj} = \frac{1}{12} \big(\frac{1}{\pi_j}+\frac{1}{\pi_k}\big)$. 
With the unadjusted $U_{jk}$, the Wilcoxon-Mann-Whitney test rejects $H_0$ when 
\begin{equation}\label{test2}
\sqrt{n} | U_{jk}- \textstyle {\frac{1}{2}} | > z_{\alpha /2} 
\sqrt{\frac{1}{12} \big( \frac{1}{\pi_j} + \frac{1}{\pi_k}\big)} , 
\end{equation} 
where $z_{\alpha/2}$ is the $1-\alpha/2$ quantile of $N(0,1)$ and $\alpha \in (0,0.5)$ is a given level of significance. Under simple randomization, the unadjusted Wilcoxon-Mann-Whitney test  given by (\ref{test2}) has asymptotic significance level $\alpha$, as we discussed in elaboration (ii) of Section 2.3. The R package wilcox.test() uses 
 (\ref{test2}) with a slight modification, i.e., $\frac{1}{\pi_j} + \frac{1}{\pi_k}$ is replaced by
$\frac{n}{n_j} + \frac{n}{n_k}+ \frac{n}{n_jn_k}$, where $\frac{n}{n_jn_k}$ is a 
continuity correction.

When covariate-adaptive randomization is applied, however, the asymptotic significance level of unadjusted Wilcoxon-Mann-Whitney test in (\ref{test2}) 
varies with randomization scheme and is not available for Pocock-Simon's minimization. 

From Theorem 1, the asymptotic variance of $U_{jk}^{ \rm C}$ is 
$\tau_{jk}+\tau_{kj}- \zeta_{jk}$. Under $H_0: F_j = F_k$, $\tau_{jk}+\tau_{kj} = \frac{1}{12} \big(\frac{1}{\pi_j}+\frac{1}{\pi_k}\big)$ and 
$\zeta_{jk}= \bbeta^\top \bSigma \bbeta \big(\frac{1}{\pi_j}+\frac{1}{\pi_k}\big)$, since $\bbeta_j = \bbeta_k = \bbeta$ not depending on $j$ and $k$. 
Let $\hat \bbeta = ( \pi_j \hat\bbeta_j + \pi_k \hat\bbeta_k)/(\pi_j+\pi_k)$, where $\hat \bbeta_j$ and $\hat \bbeta_k$ are given in (\ref{CA}).
Then, based on covariate adjusted $U_{jk}^{ \rm C}$, we propose the adjusted Wilcoxon-Mann-Whitney test that rejects $H_0$ when 
\begin{equation}\label{test1} 
\sqrt{n} | U_{jk}^{ \rm C} - \textstyle {\frac{1}{2}} | > z_{\alpha /2} 
\sqrt{ \big(\frac{1}{12} - \hat\bbeta^\top \hat\bSigma \hat\bbeta \, \big) \big(\frac{1}{\pi_j}+\frac{1}{\pi_k}\big) } , 
\end{equation}
which has asymptotic significance level $\alpha$
regardless of which covariate-adaptive randomization scheme satisfying (D) is used. 
Our covariate adjustment  widens the application scope of  Wilcoxon-Mann-Whitney test (to situations where covariate-adaptive randomization is applied). 

\subsection{Comparison of tests by Pitman's ARE}

The following result shows that, under simple randomization, adjusted Wilcoxon-Mann-Whitney test (\ref{test1})  is more efficient than unadjusted Wilcoxon-Mann-Whitney test (\ref{test2}), in terms of Pitman's asymptotic relative efficiency (ARE) \citep[pages 316-318]{Serfling}. The proof is in the Appendix. \vspace{4mm}

\noindent{\bf Theorem 2}. 
Consider  the null hypothesis $H_0\!: F_j = F_k$. Under simple randomization and 
the contiguous alternative hypothesis 
 with $F_j$ having variance $\sigma_j^2$,  continuous density $f_j$,  and $F_k(y) = F_j(y - \gamma n^{-1/2})$ for a constant $\gamma \neq 0$,  \vspace{-2mm}
\begin{enumerate}
	\item[(i)] the ARE of unadjusted  Wilcoxon-Mann-Whitney test (\ref{test2}) relative to the two sample t-test  is $ 12 \sigma_j^2 \{ \int f_j^2(y) dy \}^2$,  as $n \to \infty$; 
	\item[(ii)] the ARE of adjusted Wilcoxon-Mann-Whitney test  (\ref{test1}) 
	relative to unadjusted Wilcoxon-Mann-Whitney test  (\ref{test2}) 
	is $ 1/ (1- 12 \bbeta^\top \bSigma \bbeta) \geq 1$ with equality holds if and only if $\bbeta =0$, where $\bbeta$ is the limit of $\bbeta_j $ and $\bbeta_k$ as $n\to \infty$;
	\item[(iii)]  the ARE of adjusted Wilcoxon-Mann-Whitney test  (\ref{test1}) 
	relative to  the two sample t-test  is $ 12 \sigma_j^2 \{ \int f_j^2(y) dy \}^2/ (1- 12 \bbeta^\top \bSigma \bbeta) $. 
\end{enumerate}

Note that we do not make the ARE comparison under covariate-adaptive randomization, 
because unadjusted Wilcoxon-Mann-Whitney test (\ref{test2}) or the two sample t-test may not have asymptotic level $\alpha$ as we discussed in Sections 2.3 and 3.1. Our simulation results in Section 4 show that unadjusted Wilcoxon-Mann-Whitney test (\ref{test2}) or the two sample t-test are conservative under covariate-adaptive randomization and has power lower than that of adjusted Wilcoxon-Mann-Whitney test (\ref{test1}).
  
The efficiency of unadjusted Wilcoxon-Mann-Whitney test compared with two sample t-test is quite high since ${\rm ARE} =  0.955$ when $F_j$ is normal (in which case the use of t-test is justified), 
${\rm ARE} = 1$  when $F_j$ is uniform on the interval [0,1], ${\rm ARE} = 1.5$ when $F_j$ is double-exponential, and ARE is bounded below by 0.864 for any $F_j$ \citep{Hodges}.  
It follows from the lower bound for $ 12 \sigma_j^2  \{ \int f_j^2(y) dy \}^2$ and Theorem 2(iii) that, 
if $ 1- 12 \bbeta^\top \bSigma \bbeta< 0.864$, then adjusted Wilcoxon-Mann-Whitney test (\ref{test1}) is more powerful than the two sample t-test (without covariate adjustment) for any $F_j$ (including the case where $F_j$ is normal). 
These results are confirmed in our simulation presented in Section 4. 

\subsection{Confidence interval}

The parameter $\theta_{jk}$ measures the difference between $F_j$ and $F_k$ when $H_0$ is rejected. 
We may  assess the treatment effect by setting a confidence interval for $\theta_{jk}$. This requires a consistent estimator of
$\tau_{jk}+\tau_{kj}-\zeta_{jk}$ in Theorem 1 regardless of whether $H_0$ holds or not. While $\zeta_{jk}$ in Theorem 1 can be consistently estimated by $\hat\zeta_{jk}$ being $\zeta_{jk}$ with $\bSigma$, $\bbeta_j$ and $\bbeta_k$ substituted by $\hat\bSigma$, $\hat\bbeta_j$ and $\hat\bbeta_k$, respectively, 
regardless of whether $H_0$ holds and which covariate-adaptive randomization scheme is used, 
the remaining  term $\tau_{jk}+ \tau_{kj}$, which
is  not equal to $\frac{1}{12} ( \frac{1}{\pi_j} + \frac{1}{\pi_k})$ when $H_0$ does not hold,  has to be estimated. Since 
$$
{\rm Var} \{ F_j(Y_{k})\} = E \{ F_j^2 (Y_{k}) \}- 
[E \{ F_j(Y_{k}) \}]^2=   E \{  F_j^2 (Y_{k}) \} - \theta_{jk}^2 , 
$$
we can consistently estimate $\tau_{kj}$   by 

$$
\hat\tau_{kj}  = \frac{1}{\pi_k} \bigg[ \frac{1}{n_k} \sum_{i':A_{i'} =k } 
\bigg\{ \frac{1}{n_j} \sum_{i:A_{i}=j } I(Y_{ij} \leq Y_{i'k}) \bigg\}^2  - U_{jk}^2\bigg].
$$
Similarly, a consistent estimator of $\tau_{jk}$ is  
$$
\hat\tau_{jk}  = \frac{1}{\pi_j } \bigg[ \frac{1}{n_j} \sum_{i:A_i =j } 
\bigg\{ \frac{1}{n_k} \sum_{i':A_{i'}=k } I(Y_{ij}\leq Y_{i'k}  )\bigg\}^2
- U_{jk}^2\bigg].  $$
Hence, we estimate $\tau_{jk}+\tau_{kj}-\zeta_{jk}$ in Theorem 1 by
$\hat\tau_{jk}+ \hat\tau_{kj} - \hat \zeta_{jk} $,  
which is consistent regardless of whether $H_0$ holds or not and which covariate-adaptive randomization is used. 
A large sample level $1-\alpha$ confidence interval for $\theta_{jk}$ based on calibrated $U_{jk}^{ \rm C}$ and Theorem 1 has two end points 
$U_{jk}^{ \rm C} \pm z_{\alpha/2}  \sqrt{( \hat\tau_{jk}+ \hat\tau_{kj} - \hat \zeta_{jk} )/n}$. Without covariate adjustment, 
one can use a large sample level $1-\alpha$ confidence interval with two end points 
$U_{jk} \pm z_{\alpha/2} \sqrt{ ( \hat\tau_{jk} + \hat\tau_{kj} )/n}$, valid only under simple randomization.

\section{Simulation}

We consider a simulation study to check the finite sample performance of  unadjusted $U_{jk}$ in (\ref{wil}),  covariate adjusted $U_{jk}^{\rm C}$ in (\ref{CA}),  $\bar Y_j - \bar Y_k$ related to two sample t-test, 
and related tests for $H_0\!: F_j = F_k$ with $\alpha = 0.05$, where  $\bar Y_j$ and $\bar Y_k$ are respectively the sample means for outcomes 
in treatment groups $j$ and $k$. The two sample t-test does not adjust for covariate; it is included in our comparison to show that the better way to improve the unadjusted Wilcoxon-Mann-Whitney test is using covariate adjustment rather than t-test. 
A thorough discussion of improving t-tests using covariate adjustment is given in \cite{Ye2021better}. 
 
We generate a random sample of outcomes and covariates. 
The 2-dimensional covariate $\bX \sim $ the bivariate normal distribution with zero means, unit variances, and correlation coefficient $0.3$. 
There are a total of $J=4$ treatments with  $\pi_j =0.25$ for all $j$, but we only focus on treatments $j=1$ and $k=2$ for estimation and testing. 
To generate treatment assignments $A_i$'s, we employ simple randomization or covariate-adaptive randomization with stratified permuted block of size 8, in which strata are four categories of the first component of $\bX$ discretized with equal probabilities. 

Given treatment $A$, 
the outcome $Y_A \mid \bX$, i.e., $Y_A$ conditioned on covariate $\bX$, is generated in  the following two different cases.\vspace{-2mm}
\begin{enumerate}
	\item Normal outcome: $Y_A \mid \bX \sim $ the normal distribution with mean $ a (A-1) + (0.3,0.3)^\top \bX$ and variance  0.25, 
	where $a = 0, 0.1, 0.2$, or $0.3$. 
	\item Double-exponential outcome: 
	$Y_A \mid \bX\sim $ the double-exponential  distribution with mean  $a(A-1)+ (0.3,0.3)^\top \bX$ and variance $0.5$, where 
	$a = 0, 0.1, 0.2$, or $0.3$.
\end{enumerate}
In any case, the null hypothesis $H_0$ holds when $a=0$ and the alternative $H_1$ holds	when $a \neq 0$.

We consider  sample sizes $n =200$, 400, and 600. All observed covariates are used for covariate adjustment. The simulation results based on 5,000 replications are given in Table~\ref{t1} for normal outcome and Table~\ref{t2} for double-exponential outcome. The results include the average bias (AB), standard deviation (SD), 
average of estimated standard deviation (SE), coverage probability (CP) of the 95\% asymptotic confidence interval, type I error probability (P when $a =0$), and power (P when $a \neq 0$). 

The following is a summary of the results in Tables \ref{t1} and \ref{t2}.

\begin{enumerate}
\item The type I errors (P when $a=0$) are close to 5\% for all tests under simple randomization, but are much smaller than 5\% for t-test and unadjusted Wilcoxon-Mann-Whitney test under covariate-adaptive randomization, 
showing that they are conservative as we discussed in Section 3. This conservativeness also affects the coverage probability of confidence intervals using sample means and unadjusted $U_{jk}$. 
\item In terms of SD, the adjusted $U_{jk}^{\rm C}$ in (\ref{CA}) is much better than the unadjusted $U_{jk}$ in (\ref{wil}).
\item In terms of power (P when $a \neq 0$), the two sample t-test compared with unadjusted Wilcoxon-Mann-Whitney test (\ref{test2}) 
is slightly better for normal outcome (Table~\ref{t1}), but worse for 
double-exponential outcome (Table~\ref{t2}). Covariate  adjusted Wilcoxon-Mann-Whitney test (\ref{test1}) 
is much more powerful than the two sample t-test and unadjusted Wilcoxon-Mann-Whitney test (\ref{test2}) (both of which have no covariate adjustment), 
regardless of whether outcome is normal or double-exponential or whether covariate-adaptive randomization is applied. This confirms our asymptotic theory in Section 3.3. 
\item The proposed variance estimator (SE) works well and the coverage probabilities are close to the targeted 95\%. 
\end{enumerate}

\begin{table}[htbp!]
	\centering
	\caption{Simulation results based on 5,000 replications for normal  outcome:  AB = average of bias, SD = standard deviation, SE = average of estimated SD, CP = coverage probability of 95\% asymptotic confidence interval, P = type I error probability when $a=0$ and P = power when $a \neq 0$.}\vspace{2mm}
	\label{t1}
	\begin{tabular}{cccccccccccccc}
		\toprule
		&  &  & \multicolumn{5}{c}{Simple randomization} && \multicolumn{5}{c}{Stratified permuted block} \\  \cline{4-8} \cline{10-14}
		\vspace{-2.5mm}	$~$ & \\ 
		$a$ & $n$ & Estimator & AB   & SD & SE & CP & P  &  & AB & SD & SE & CP &P  \\ \cline{1-14}
		\vspace{-2mm}	$~$ & \\ 
		0 & 200 & $\bar Y_j - \bar Y_k$ & -0.002 & 0.141 & 0.142 & 0.948 & 0.052 && 0.002 & 0.120 & 0.141 & 0.974 & 0.026 \\
		& & $U_{jk}$ & -0.001 & 0.059 & 0.058 & 0.938 & 0.057 && 0.001 & 0.050 & 0.058 & 0.971 & 0.027 \\
		& & $U_{jk}^{\rm C}$  & 0.000 & 0.046 & 0.045 & 0.938 & 0.052 && 0.001 & 0.044 & 0.044 & 0.945 & 0.048 \\
		& 400 & $\bar Y_j - \bar Y_k$& -0.003 & 0.101 & 0.099 & 0.944 & 0.056 && 0.002 & 0.085 & 0.099 & 0.975 & 0.025 \\
		& & $U_{jk}$ & -0.001 & 0.042 & 0.041 & 0.944 & 0.053 && 0.001 & 0.036 & 0.041 & 0.973 & 0.025 \\
		& & $U_{jk}^{\rm C}$ & 0.000 & 0.031 & 0.031 & 0.942 & 0.055 && 0.001 & 0.031 & 0.031 & 0.948 & 0.050 \\
		& 600 & $\bar Y_j - \bar Y_k$& 0.000 & 0.080 & 0.081 & 0.954 & 0.046 && -0.001 & 0.069 & 0.081 & 0.975 & 0.025 \\
		& & $U_{jk}$ & 0.000 & 0.033 & 0.033 & 0.951 & 0.047 && 0.000 & 0.029 & 0.033 & 0.974 & 0.025 \\
		& & $U_{jk}^{\rm C}$ & 0.000 & 0.025 & 0.025 & 0.947 & 0.051 && 0.000 & 0.025 & 0.024 & 0.952 & 0.048 \\
		\vspace{-2mm}	$~$ & \\ 
		0.1& 200 &$\bar Y_j - \bar Y_k$  & -0.002 & 0.141 & 0.142 & 0.948 & 0.112 && 0.002 & 0.120 & 0.141 & 0.974 & 0.076 \\
		& & $U_{jk}$ & 0.000 & 0.059 & 0.058 & 0.938 & 0.112 && 0.002 & 0.050 & 0.058 & 0.969 & 0.078 \\
		& & $U_{jk}^{\rm C}$  & 0.000 & 0.046 & 0.044 & 0.936 & 0.155 && 0.002 & 0.044 & 0.044 & 0.946 & 0.152 \\
		& 400 &$\bar Y_j - \bar Y_k$ & -0.003 & 0.101 & 0.099 & 0.944 & 0.175 && 0.002 & 0.085 & 0.099 & 0.975 & 0.141 \\
		& & $U_{jk}$ & -0.001 & 0.042 & 0.041 & 0.944 & 0.168 && 0.002 & 0.035 & 0.041 & 0.974 & 0.138 \\
		& & $U_{jk}^{\rm C}$ & 0.000 & 0.031 & 0.030 & 0.942 & 0.254 && 0.001 & 0.031 & 0.030 & 0.948 & 0.265 \\
		& 600 & $\bar Y_j - \bar Y_k$ & 0.000 & 0.080 & 0.081 & 0.954 & 0.236 && -0.001 & 0.069 & 0.081 & 0.975 & 0.193 \\
		& & $U_{jk}$ & 0.000 & 0.033 & 0.033 & 0.951 & 0.229 && 0.000 & 0.029 & 0.033 & 0.975 & 0.184 \\
		& & $U_{jk}^{\rm C}$ & 0.000 & 0.025 & 0.024 & 0.946 & 0.374 && 0.000 & 0.025 & 0.024 & 0.949 & 0.367 \\
		\vspace{-2mm}	$~$ & \\ 
		0.2 & 200 & $\bar Y_j - \bar Y_k$  & -0.002 & 0.141 & 0.142 & 0.948 & 0.286 && 0.002 & 0.120 & 0.141 & 0.974 & 0.266 \\
		& & $U_{jk}$ & 0.000 & 0.058 & 0.057 & 0.938 & 0.281 && 0.001 & 0.049 & 0.057 & 0.968 & 0.260 \\
		& & $U_{jk}^{\rm C}$  & 0.001 & 0.045 & 0.043 & 0.936 & 0.438 && 0.002 & 0.043 & 0.043 & 0.945 & 0.448 \\
		& 400 & $\bar Y_j - \bar Y_k$ & -0.003 & 0.101 & 0.099 & 0.944 & 0.497 && 0.002 & 0.085 & 0.099 & 0.975 & 0.537 \\
		& & $U_{jk}$ & 0.000 & 0.041 & 0.040 & 0.943 & 0.485 && 0.001 & 0.035 & 0.040 & 0.975 & 0.516 \\
		& & $U_{jk}^{\rm C}$ & 0.001 & 0.031 & 0.030 & 0.942 & 0.735 && 0.001 & 0.030 & 0.030 & 0.949 & 0.759 \\
		& 600 & $\bar Y_j - \bar Y_k$  & 0.000 & 0.080 & 0.081 & 0.954 & 0.700 && -0.001 & 0.069 & 0.081 & 0.975 & 0.726 \\
		& & $U_{jk}$ & 0.000 & 0.032 & 0.033 & 0.949 & 0.679 && 0.000 & 0.028 & 0.033 & 0.973 & 0.699 \\
		& & $U_{jk}^{\rm C}$ & 0.000 & 0.024 & 0.024 & 0.947 & 0.904 && 0.000 & 0.024 & 0.024 & 0.950 & 0.906 \\
		\vspace{-2mm}	$~$ & \\ 
		0.3 & 200 & $\bar Y_j - \bar Y_k$ & -0.002 & 0.141 & 0.142 & 0.948 & 0.556 && 0.002 & 0.120 & 0.141 & 0.974 & 0.585 \\
		& & $U_{jk}$ & -0.001 & 0.057 & 0.056 & 0.940 & 0.538 && 0.001 & 0.048 & 0.056 & 0.968 & 0.564 \\
		& & $U_{jk}^{\rm C}$  & 0.001 & 0.044 & 0.043 & 0.934 & 0.757 && 0.002 & 0.042 & 0.042 & 0.946 & 0.791 \\
		& 400 & $\bar Y_j - \bar Y_k$ & -0.003 & 0.101 & 0.099 & 0.944 & 0.842 && 0.002 & 0.085 & 0.099 & 0.975 & 0.901 \\
		& & $U_{jk}$ & 0.000 & 0.041 & 0.039 & 0.944 & 0.827 && 0.001 & 0.034 & 0.039 & 0.973 & 0.887 \\
		& & $U_{jk}^{\rm C}$ & 0.001 & 0.030 & 0.030 & 0.943 & 0.975 && 0.001 & 0.030 & 0.030 & 0.945 & 0.976 \\
		& 600 & $\bar Y_j - \bar Y_k$  & 0.000 & 0.080 & 0.081 & 0.954 & 0.960 && -0.001 & 0.069 & 0.081 & 0.975 & 0.977 \\
		& & $U_{jk}$ & 0.001 & 0.032 & 0.032 & 0.952 & 0.956 && 0.000 & 0.028 & 0.032 & 0.973 & 0.973 \\
		& & $U_{jk}^{\rm C}$ & 0.001 & 0.024 & 0.024 & 0.946 & 0.998 && 0.000 & 0.024 & 0.024 & 0.949 & 0.999 \\
		\bottomrule
	\end{tabular}
\end{table}

\begin{table}[htbp!]
	\centering
	\caption{Simulation results based on 5,000 replications for double-exponential  outcome: AB = average of bias, SD = standard deviation, SE = average of estimated SD, CP = coverage probability of 95\% asymptotic confidence interval, P = type I error probability when $a=0$ and P = power when $a \neq 0$.} \vspace{2mm}
	\label{t2}
	\begin{tabular}{cccccccccccccc}
		\hline
		&  &  & \multicolumn{5}{c}{Simple randomization} && \multicolumn{5}{c}{Stratified permuted block} \\  \cline{4-8} \cline{10-14}
		\vspace{-2.5mm}	$~$ & \\ 
		$a$ & $n$ & Estimator & AB   & SD & SE & CP & P  &  & AB & SD & SE & CP &P  \\ \hline
		\vspace{-2mm}	$~$ & \\ 
		0 & 200 & $\bar Y_j - \bar Y_k$ & -0.003 & 0.173 & 0.174 & 0.956 & 0.044 && -0.002 & 0.155 & 0.173 & 0.973 & 0.027 \\
		& & $U_{jk}$ & -0.001 & 0.058 & 0.058 & 0.947 & 0.048 && -0.001 & 0.052 & 0.058 & 0.971 & 0.026 \\
		& & $U_{jk}^{\rm C}$  & -0.001 & 0.048 & 0.048 & 0.944 & 0.051 && -0.001 & 0.048 & 0.047 & 0.942 & 0.052 \\
		& 400 & $\bar Y_j - \bar Y_k$& 0.001 & 0.122 & 0.122 & 0.954 & 0.046 && -0.001 & 0.111 & 0.122 & 0.965 & 0.035 \\
		& & $U_{jk}$ & 0.000 & 0.041 & 0.041 & 0.947 & 0.051 && 0.000 & 0.037 & 0.041 & 0.968 & 0.030 \\
		& & $U_{jk}^{\rm C}$ & 0.000 & 0.034 & 0.033 & 0.942 & 0.054 && 0.000 & 0.033 & 0.033 & 0.952 & 0.046 \\
		& 600 & $\bar Y_j - \bar Y_k$& -0.001 & 0.098 & 0.099 & 0.949 & 0.051 && 0.000 & 0.090 & 0.099 & 0.972 & 0.028 \\
		& & $U_{jk}$ & 0.000 & 0.033 & 0.033 & 0.946 & 0.052 && 0.000 & 0.030 & 0.033 & 0.970 & 0.029 \\
		& & $U_{jk}^{\rm C}$ & 0.000 & 0.027 & 0.027 & 0.947 & 0.052 && 0.000 & 0.027 & 0.027 & 0.945 & 0.053 \\
		\vspace{-2mm}	$~$ & \\ 
		0.1& 200 &$\bar Y_j - \bar Y_k$  & -0.003 & 0.173 & 0.174 & 0.955 & 0.079 && -0.001 & 0.155 & 0.173 & 0.973 & 0.064 \\
		& & $U_{jk}$ & 0.000 & 0.058 & 0.058 & 0.949 & 0.090 && -0.001 & 0.052 & 0.058 & 0.970 & 0.065 \\
		& & $U_{jk}^{\rm C}$  & 0.000 & 0.048 & 0.047 & 0.943 & 0.112 && -0.001 & 0.048 & 0.047 & 0.941 & 0.113 \\
		& 400 &$\bar Y_j - \bar Y_k$ & 0.001 & 0.122 & 0.122 & 0.954 & 0.135 && -0.001 & 0.111 & 0.122 & 0.965 & 0.103 \\
		& & $U_{jk}$ & 0.000 & 0.041 & 0.041 & 0.946 & 0.142 && 0.000 & 0.037 & 0.041 & 0.967 & 0.111 \\
		& & $U_{jk}^{\rm C}$ & 0.001 & 0.034 & 0.033 & 0.941 & 0.192 && 0.001 & 0.033 & 0.033 & 0.951 & 0.194 \\
		& 600 & $\bar Y_j - \bar Y_k$ & -0.003 & 0.098 & 0.099 & 0.949 & 0.168 && 0.001 & 0.090 & 0.099 & 0.972 & 0.148 \\
		& & $U_{jk}$ & -0.001 & 0.033 & 0.033 & 0.947 & 0.181 && -0.001 & 0.030 & 0.033 & 0.970 & 0.156 \\
		& & $U_{jk}^{\rm C}$ & -0.001 & 0.027 & 0.027 & 0.947 & 0.259 && -0.001 & 0.027 & 0.027 & 0.943 & 0.264 \\
		\vspace{-2mm}	$~$ & \\ 
		0.2 & 200 & $\bar Y_j - \bar Y_k$  & -0.004 & 0.173 & 0.174 & 0.955 & 0.215 && -0.005 & 0.155 & 0.173 & 0.972 & 0.185 \\
		& & $U_{jk}$ & -0.001 & 0.057 & 0.057 & 0.949 & 0.231 && -0.002 & 0.051 & 0.057 & 0.970 & 0.202 \\
		& & $U_{jk}^{\rm C}$  & 0.000 & 0.048 & 0.047 & 0.947 & 0.309 && -0.001 & 0.048 & 0.047 & 0.938 & 0.313 \\
		& 400 & $\bar Y_j - \bar Y_k$ & 0.002 & 0.122 & 0.122 & 0.954 & 0.378 && 0.000 & 0.111 & 0.122 & 0.965 & 0.363 \\
		& & $U_{jk}$ & 0.000 & 0.041 & 0.040 & 0.947 & 0.409 && 0.000 & 0.036 & 0.040 & 0.966 & 0.399 \\
		& & $U_{jk}^{\rm C}$ & 0.000 & 0.034 & 0.033 & 0.941 & 0.571 && 0.001 & 0.033 & 0.033 & 0.951 & 0.575 \\
		& 600 & $\bar Y_j - \bar Y_k$  & -0.002 & 0.098 & 0.099 & 0.949 & 0.510 && 0.001 & 0.090 & 0.099 & 0.972 & 0.523 \\
		& & $U_{jk}$ & -0.001 & 0.033 & 0.033 & 0.948 & 0.555 && 0.000 & 0.029 & 0.033 & 0.971 & 0.575 \\
		& & $U_{jk}^{\rm C}$ & 0.000 & 0.027 & 0.027 & 0.947 & 0.747 && 0.000 & 0.027 & 0.026 & 0.944 & 0.743 \\
		\vspace{-2mm}	$~$ & \\ 
		0.3 & 200 & $\bar Y_j - \bar Y_k$ & -0.002 & 0.173 & 0.174 & 0.956 & 0.415 && -0.002 & 0.155 & 0.173 & 0.973 & 0.402 \\
		& & $U_{jk}$ & 0.000 & 0.057 & 0.057 & 0.950 & 0.442 && -0.001 & 0.051 & 0.056 & 0.968 & 0.445 \\
		& & $U_{jk}^{\rm C}$  & 0.001 & 0.047 & 0.046 & 0.945 & 0.592 && 0.000 & 0.047 & 0.046 & 0.941 & 0.593 \\
		& 400 & $\bar Y_j - \bar Y_k$ & 0.001 & 0.122 & 0.122 & 0.954 & 0.697 && 0.000 & 0.111 & 0.122 & 0.965 & 0.708 \\
		& & $U_{jk}$ & 0.000 & 0.040 & 0.040 & 0.949 & 0.739 && 0.000 & 0.036 & 0.040 & 0.967 & 0.755 \\
		& & $U_{jk}^{\rm C}$ & 0.001 & 0.033 & 0.033 & 0.942 & 0.881 && 0.001 & 0.033 & 0.032 & 0.951 & 0.888 \\
		& 600 & $\bar Y_j - \bar Y_k$  & -0.003 & 0.098 & 0.099 & 0.949 & 0.854 && 0.001 & 0.090 & 0.099 & 0.972 & 0.876 \\
		& & $U_{jk}$ & -0.001 & 0.032 & 0.033 & 0.949 & 0.882 && 0.000 & 0.029 & 0.032 & 0.971 & 0.916 \\
		& & $U_{jk}^{\rm C}$ & 0.000 & 0.027 & 0.026 & 0.947 & 0.974 && 0.000 & 0.026 & 0.026 & 0.943 & 0.974 \\
		\bottomrule					
	\end{tabular}
\end{table}

\section{Example}

We  consider a dataset from a randomized  phase 3 clinical trial sponsored by Eli Lilly and Company, for adults with type 2 diabetes who remained inadequately controlled on standard oral glucose-lowering therapy. 
Eligible participants are randomly assigned  in a 1:1:1:3 ratio to  one of three treatment dose arms and an active control group,  using stratified permuted block randomization with block size 6 to ensure balance across important baseline characteristics.  
To illustrate the proposed method, we focus on a pre-specified exploratory outcome,  the change in total cholesterol from baseline to week 52. 
We restrict the analysis to the subgroup of  $n=491$ participants with baseline body mass index (BMI) greater than $35\text{kg/m}^2$, 
$n_1=76$ randomized to treatment dose 1, $n_2=83$ to treatment dose 2, $n_3=88$ to treatment dose 3, and $n_4=244$  to control group. 
In this subgroup, the mean age is 62 years, 45\% are female,  and the median baseline total cholesterol is 154 mg/dL with an interquartile range of 130 to 183 mg/dL.

For pairwise comparisons between the control group and each treatment arm, we test three null hypotheses of interest: $H_0\!: F_j = F_4$, where $F_4$ is the outcome distribution under control group and $F_j$ is the outcome distribution  under treatment with dose  $j=1,2,3$.


We consider four testing methods. 
\begin{enumerate}
	\item[(i)] The  two sample t-test applied to raw changes in total cholesterol. Note that the two sample t-test is not for $H_0\!: F_j= F_4$ unless  raw changes in total cholesterol are normally distributed with the same variance across treatment arms. 
	\item[(ii)] The two sample t-test applied to transformed outcome, 
		log (total cholesterol at week 52) $-$ log (total cholesterol at baseline). 
		We consider this transformation because  lipid data often exhibit skewed distribution 
		and  raw changes in total cholesterol  do not appear to be normally distributed and  have large variability. 
		\item[(iii)] The unadjusted Wilcoxon-Mann-Whitney  test applied 
	to raw changes  in total cholesterol.
	\item[(iv)] The covariate adjusted Wilcoxon-Mann-Whitney  test  applied to  raw changes  in total cholesterol, with adjustment using five baseline covariates,  the age, systolic blood pressure, hemoglobin A1c, BMI, and baseline total cholesterol. 
\end{enumerate}


\begin{table}[htbp]
		\centering
	\caption{Comparison on change in total cholesterol under different treatment doses versus the active control from a completed phase 3 trial for the treatment of diabetes.}
	\vspace{3mm}
	\label{t3}
		\begin{tabular}{cccccc}
			\toprule
			&  & \multicolumn{4}{c}{Testing method}                       \\ \cline{3-6} \noalign{\medskip}  
			& & \multicolumn{2}{c}{Two sample t-test} & \multicolumn{2}{c}{Wilcoxon-Mann-Whitney test} \\
			Comparison                              &                       & raw data&  transformed data& unadjusted  & covariate adjusted \\ \hline \noalign{\medskip}  
			{treatment dose 1 vs control} & p-value               &   0.122  &   0.051      &      0.023    &     0.044     \\
			& SE                    &  4.746      &   0.028      &       0.038         &     0.034         \\
			& CI                    &  (-16.812, 2.002)   &   (-0.112, 0.000)       & (0.339, 0.489) &   (0.365, 0.498)       \\ \noalign{\medskip}  
			
			{treatment dose 2 vs control} & p-value               &   0.089  &   0.081      &      0.063    &     0.022     \\
			& SE                    &  4.486      &   0.029      &       0.038         &     0.031         \\
			& CI                    &  (-16.557, 1.201)   &   (-0.106, 0.006)       & (0.358, 0.506) &   (0.367, 0.486)       \\ \noalign{\medskip}  
			
			{treatment dose 3 vs control} & p-value               &   0.112  &   0.039      &      0.079    &     0.045     \\
			& SE                    &  4.391      &   0.027      &       0.037         &     0.034         \\
			& CI                    &  (-15.719, 1.651)   &   (-0.110, -0.003)       & (0.364, 0.509) &   (0.368, 0.501)       \\ \noalign{\medskip}  
			\bottomrule \noalign{\medskip}  
			\multicolumn{6}{l}{SE: the standard error of difference of sample means (raw data or transformed) under two sample t-test} \\
			\multicolumn{6}{l}{SE: the standard error of  Wilconxon statistic (unadjusted or adjusted) under Wilcoxon-Mann-Whitney test}
	\end{tabular}
\end{table}

For each pairwise comparison, Table~\ref{t3} provides the two-sided p-value for each of the four testing methods,  the standard error (SE) of the difference of sample means from treatment $j$ and $4$ when the two sample t-test (saw data or transformed data) is used or the SE of $U_{j4}$ or $U_{j4}^{\rm C}$ when Wilcoxon-Mann-Whitney test (unadjusted or adjusted) is used, and the confidence interval (CI) related to each  testing methods.  

The empirical results in Table~\ref{t3}
demonstrate that the covariate adjusted 
$U_{j4}^{\rm C}$ has smaller SE than the unadjusted $U_{j4}$, which is consistent with our theory and simulation results. When 5\% is considered as the significance level, the covariate adjusted  Wilcoxon-Mann-Whitney test rejects all three null hypotheses in pairwise comparisons, whereas the unadjusted Wilcoxon-Mann-Whitney test cannot reject when $j=2$ and $j=3$, indicating its inefficiency. 
The two sample t-test applied to raw data in this example is apparently not appropriate, since it cannot reject any null hypothesis and its related confidence interval is uselessly too wide. 

Although the two sample t-test applied to log-transformed outcomes  is better than that based on raw data, its performance is still not as good as the Wilcoxon-Mann-Whitney test with covariate adjustment.  Furthermore, the interpretation of confidence intervals based on transformed outcomes  is not straightforward, since 
 the confidence interval  on the log scale does not translate directly back to the original outcome scale and may lead to incorrect interpretations, due to the fact that  $E( \log Y) \neq \log E(Y)$ (geometric mean versus arithmetic mean).
 By contrast, the confidence intervals based on Wilcoxon two sample statistics, adjusted or unadjusted, are on the original scale without any transformation needed to meet distributional requirement.  
 








\bibliographystyle{apalike}
\bibliography{reference}



\appendix

\section{Proof of Theorem 1}

From page 381 of \cite{Jiang2010}, 
$$ \sqrt{n} (U_{jk} - \theta_{jk}) = \sqrt{n}(\bar W_{jk} + \bar W_{kj})+ o_p(1), $$
where
$$ 
\bar W_{jk} =  \frac{1}{n_j} \sum_{i:A_i=j} \{ 1- F_k(Y_{ij}) - \theta_{jk} \} , \qquad \bar W_{kj} = \frac{1}{n_k} \sum_{i:A_i=k} \{ F_j(Y_{ik}) - \theta_{jk} \} ,
$$
and $o_p(1)$ denotes a term converges to 0 in probability.
Consequently, 
\begin{align*}
\sqrt{n} (U_{jk}^{ \rm C} - \theta_{jk}) & = \sqrt{n} (\bar W_{jk} + \bar W_{kj})
+  (\bar{\bX}_{j}-\bar{\bX})^\top \hat \bbeta_j
- (\bar{\bX}_{k}-\bar{\bX})^\top \hat \bbeta_k 
+ o_p(1) \\ 
& = \sqrt{n} (\bar W_{jk} + \bar W_{kj})
+  (\bar{\bX}_{j}-\bar{\bX})^\top  \bbeta_j
- (\bar{\bX}_{k}-\bar{\bX})^\top  \bbeta_k 
+ o_p(1) .
\end{align*}

\noindent
The rest of proof follows the same argument in the proofs of Corollary 1 and Theorem 2 in \cite{Ye2021better}, since $\bar W_{jk}$ and $\bar W_{kj}$ are types of sample means with outcomes in treatment groups $j$ and $k$, respectively. 

\section{Proof of Theorem 2}

(i) Under the contiguous alternative hypothesis
with $F_j$ having mean 0, variance $\sigma_j^2$, and continuous density $f_j$,  and $F_k(y) = F_j(y - \gamma n^{-1/2})$ with a constant $\gamma \neq 0$,  
$ \sqrt{n} (U_{jk} - \theta_{jk}) $
converges in distribution to the normal distribution with mean 0 and variance $\frac{1}{12} \big( \frac{1}{\pi_j} + \frac{1}{\pi_k}\big)$. Let $\mu = \gamma n^{-1/2}$. Then
$$
\theta_{jk} =  E(U_{jk}) =   \int \{ 1-F_k(y)\}dF_j(y)    =  \int \{ 1-F_j(y-\mu )\}f_j(y) dy  
$$
and, under the assumed conditions, 
$$
\frac{d \theta_{jk}}{d \mu} = \int f_j(y-\mu )f_j(y) dy \to \int f_j^2(y) dy . 
$$
The two sample t-test is based on the fact that 
$\sqrt{n}  ( \bar Y_k - \bar Y_j - \mu )$ converges in distribution to the normal distribution with mean 0 and variance
 $\sigma_j^2\big( \frac{1}{\pi_j} + \frac{1}{\pi_k}\big)$. 
 Hence, the ARE of unadjusted Wilcoxon-Mann-Whitney test (\ref{test2}) relative to two sample t-test is equal to
 \citep[pages 316-318]{Serfling}
 $$
 {\rm ARE} = \lim_{n\to \infty} \frac{\sigma_j^2\big( \frac{1}{\pi_j} + \frac{1}{\pi_k}\big)}{\big(\frac{d \mu}{ d \mu}\big)^2 } \frac{\big(\frac{d \theta_{jk}}{ d \mu}\big)^2 }{\frac{1}{12} \big( \frac{1}{\pi_j} + \frac{1}{\pi_k}\big)} = 12 \sigma_j^2 \bigg\{ \int f_j^2(y) dy \bigg\}^2 
 $$
(ii)  The ARE of adjusted Wilcoxon-Mann-Whitney test (\ref{test1}) relative to 
 unadjusted Wilcoxon-Mann-Whitney test (\ref{test2}) can be similarly obtained as
 $$
 {\rm ARE} = \lim_{n\to \infty} \frac{\frac{1}{12} \big( \frac{1}{\pi_j} + \frac{1}{\pi_k}\big)}{\big(\frac{d \theta_{jk}}{ d \mu}\big)^2  } \frac{\big(\frac{d \theta_{jk}}{ d \mu}\big)^2 }{\frac{1}{12} \big( \frac{1}{\pi_j} + \frac{1}{\pi_k}\big) - \zeta_{jk}} = 
 \frac{1}{ 1- 12 \bbeta^\top \bSigma \bbeta}. 
 $$
(iii) The ARE in (iii) is equal to the ARE in (i) multiplying the ARE (ii).

\end{document}